\documentstyle[12pt]{article}
\input{psfig.tex}
\def\be{\begin{equation}}
\def\ee{\end{equation}}
\def\bea{\begin{eqnarray}}
\def\eea{\end{eqnarray}}

\begin{document}
\pagestyle{empty}
\begin{flushright}
{BROWN-HET-1216}\\
{February 2000}
\end{flushright}
\vskip .15in

\begin{center}
LUMPS AND P-BRANES IN OPEN STRING FIELD THEORY\\
\vskip .15in
R. DE MELLO KOCH\\
{\it  Physics Department, University of Witwatersrand}\\
{\it Johannesburg, South Africa} \\
{\it and}\\ 
 A. JEVICKI, M. MIHAILESCU, R. TATAR\\
{\it Department of Physics, Brown University}\\
{\it  Providence, Rhode Island  02912  USA}
\vskip .50in
{\bf ABSTRACT}
\end{center}

We describe numerical methods for constructing lump solutions in open 
string field theory. According to  Sen, these lumps represent 
lower dimensional Dp-Branes and numerical evaluation of their energy can be 
compared with the expected value for the tension.  We take particular care of 
all higher derivative terms inherent in 
Witten's version of open string field theory.  The importance of these terms
for off shell phenomena is argued in the 
text.  Detailed numerical calculations done for the case of general $p$ brane 
show very good agreement with Sen's conjectured value. This gives  credence to the 
conjecture itself and establishes further the usefulness of Witten's 
version of SFT .   

\vskip .10in
\newpage
\setcounter{page}{1}
\pagestyle{plain}

\section{Introduction}
\medskip
Since its introduction, string field theory held out the promise for 
nonperturbative studies of string theory.  Recently Sen \cite{one, two} has argued that open 
bosonic (string) field theory describes the dynamics of 
$D\bar{D}$ system with the tachyon providing the instability inherent in such 
pair.  Tachyon condensation also describes the decay of a single unstable 
$D$-brane. In addition,
according to \cite{two} the kink and lump solutions of such field theory lead to 
lower dimensional branes.  In recent work Sen and Zweibach \cite{three}  
studied in detail tachyon condensation in  26D open string theory.  This follows 
the earlier, pioneering work of Samuel and Kostelecky who were the first to 
consider the vacuum structure of string field theory \cite{four, five}.  They used a level 
truncation scheme to generate an approximation for the tachyon effective 
potential.  Following this scheme, results were found  \cite{three} that  
 show great agreement in numerical values with the expected exact results.
Similar results were recently obtained for superstring field theory \cite{six, seven}. 
Impressive high level studies appeared in \cite{eight, nine}.

It is equally important to give a construction of 
non-constant kink and lump-like solutions.  One can expect that it is for 
these 
that the stringy effects present in string field theory might play the most 
important role.  One of the characteristic features present in construction  of 
Witten's
version of the theory \cite{ten, eleven, twelve, thirteen} is the appearance in the interaction of terms 
exponential in derivatives.   These terms can be moved from the interaction to give a nontrivial 
kinetic term. They
have  frustrated early attempts for construction of nonperturbative soliton 
(or instanton) solutions of the theory.  

In the present work,  we consider this  
problem  in 
numerical terms.  Concentrating on the lump of open string theory, we 
develop methods for its  numerical solution.  In this we keep the nontrivial 
exponential terms characteristic of the string vertex interaction and 
simultaneously perform a level truncation.  This, we argue is  to provide a 
very good approximation to the exact result.

The content of the paper is as follows.  After a short description of open SFT, 
we discuss some features relevant to the present work.  We explain (based on 
earlier observations \cite{fourteen}) how and why the approach of keeping higher derivatives
 and 
simultaneous level truncation  holds the promise for a good 
approximation.  We then proceed to the numerical work. 

While this work was in progress, there appeared the work of Ref.\cite{fifteen} which 
considers 
the problem in its field theory limit.
\medskip
\section{Open String Field Theory}
\medskip

We begin by describing some features of  open string field theory which 
are of relevance to the investigation that follows. One has the cubic 
action:
\be
S = \langle A \vert Q \vert A \rangle + {g\over 3} \langle V_3 \vert\vert A 
\rangle \vert A \rangle \vert A \rangle
\ee
with Q being the first quantized BRST operator. The geometric, three string
interaction is realized in the Hilbert space  by the 
vertex
\be
\langle V_3 \vert = \langle 0 \vert \exp \{\sum{\alpha_{n}}^{r} N_{nm}^{rs} 
\alpha_{m}^{s} + \alpha ' \ln \gamma \sum_{r=1}^{3} \partial_{r}^{2}\}
\ee
Here the Neumann coefficient $N^{rs}$ are determined in terms of appropriate 
conformal mapping ,their explicit values are determined in \cite{ten, eleven}.

One of the main properties of the three-string vertex is an  
explicit appearance of higher derivative terms.  
They come in  exponential form acting on each string field
\be
\vert A \rangle\rightarrow \exp \left( \alpha ' \ln  \gamma \partial_{\mu} \partial^{\mu} 
\right)\vert A\rangle 
\ee
with the constant
\be
\gamma = {3\sqrt{3}\over 4}
\ee
The exponential terms are not relevant in studies of vacuum structure but
they can have a nontrivial effect in any other nonzero momentum process. Concerning a systematic 
approximation or
expansion scheme one notes the following. 
The masses of tachyon (and higher mass fields) are proportional 
to $1 /\alpha ' \ $. In general $\alpha '$ serves as a scale ,it  can be scaled 
out in front of the SFT action. 
 At a nonperturbative level, there is no small free parameter  and 
no systematic expansion.  Since  one is not able(at present)  to  
solve the theory in exact terms, one relies on seemingly and hoc 
approximations. Such is the process of level truncation. In order to
understand more clearly the procedure involved and  the relative relevance of particular 
terms, let us recall the original 
argument given for the level truncation in an unpublished work of ref.\cite{fourteen}.  
Considering an approximate 
calculation of a nonzero momentum amplitude, for example for four tachyons one starts from: 
\be
A_s = \int_0^1 dx \, x^{-s/2-2} \langle V_{34} (\bar{3'} ) \vert b_0 x^R \vert 
V_{12} (3')\rangle ,
\ee
representing  the $s$-chanel Feynman diagram.

Using the fact that $x^R \alpha_n x^{-R} = x^n \alpha_n$, as well as similar 
results for $b's$ and $c's$, we have
\be
A_s = \int_0^1 dx \, x^{-s/2-2} \langle V_{34} (\dot{3} ) b_0 \vert V_{12} 
(\dot{3} ) \rangle .
\ee

The dot on $V$ indicates that $a_{-n},  b_{-n},  c_{-n}$ have been replaced by 
$ x^n\linebreak
 \alpha_{-n},    x^n  b_{-n},    x^n c_{-n} $.  Expanding in levels 
corresponds  to expanding  in powers of $x=e^{-r}$.  
With the use  the appropriate Neumann coefficients after a straightforward 
algebra one has
\be
A_s = \int_0^1 dx \, x^{-s/2-2} \left( 1 - {11^2\over 3^6} x^2 + \cdots \right) 
e^{E(x)} \, .
\ee

The term in the exponent has the expansion

\bea E(x) = - {s\over 2} \ln \gamma & - & \left( {s\over 2 }+ 2\right) \left( - 
{2^3\over 3^3} x + {2^2 \cdot 19\over 3^6} x^2 + \cdots \right) - \nonumber \\
& - & \left( {i\over 2} + 2\right) \left( - {2^4\over 3^3} x - {2^6 \cdot 
7^2\over 3^{10}} x^3 + \cdots \right) - \nonumber \\
& - & \left( {2^4\over 3^3} x + {2^3\over 3^6} x^2 + \cdots \right) + \left( 
{26\cdot 5^2\over 2\cdot 3^6} x^2 + \cdots \right) \, .
\eea

This result can easily be rearranged into a form
\be
A_s = \int_0^{z(1)} dz \, z^{-s/2-2} (1-z)^{-t/2-2}
\ee
where
\be
z(x) = {1\over \gamma^2} \, x \left( 1 - {2^3\over 3^3} x + {2^2\over 3^3} \, 
x^2 + \dots \right) \, .
\ee

For agreement with the exact result one would  need to have z(1)=0.5 (the s and t -chanel diagrams
are to cover the full range (0,1).To the present 
order in the level expansion  we have $z (1) \approx 0.50480$ which is indeed 
very close to the exact value. A more significant observation is the fact that 
the main effect is contained in the $\, \gamma^{-2} \,$ factor present in the above 
expression . That term itself gives $z(1) \approx 0.6 $. If we look up the 
origin of this factor, we see that it comes directly from the exponential higher derivative
operator present in the vertex: since the intermediate states in the 
four point amplitude calculation are not on shell the exponential terms
contribute giving the corresponding $s$ dependence. 

In this nonzero momentum example, we conclude that it is advantageous to keep 
the higher derivatives exactly and that this followed by a level truncation is likely to 
give a good overall approximation. Naturaly one still expects this expansion to only
be good for certain range of momenta.

Consider then the string field theory with the tachyon (level 0), but with the higher derivative terms 
kept exactly.  
The action  evaluated originally in \cite{eleven}  reads 

\be
{\cal L} = {1\over 2} \left( \partial_{\mu} T\right)^2 + {1\over 2\alpha '} \, 
T^2 - {g\over 3} \gamma^3 \tilde{T}^3  \,\, {\rm with} \,\, \tilde{T}= e^{\ln\gamma\partial^{2}}T 
\ee

The first sign that the presence of the  exponential term in the cubic interaction 
profoundly influences the nature of the problem is seen in attempting to evaluate the 
asymptotics of a possible static solution.  In ordinary field theories, 
kink and lump solution can be asymptoticaly characterized  as 

\be
\phi (x) \sim \phi_0 + a_1 e^{-mx}
\ee
where $\phi_0$ is the constant vacuum solution and $m$ is the physical mass of 
the scalar field.  Based on this, one can write a systematic expansion for the 
lump
\be
\phi (x) = \sum_{m=0}^{\infty} a_n \, e^{-nmx}
\ee
with the classical equations providing a recursion formula for coefficients $a_n$.  
The exponential decay  $e^{-nmx}$ has a physical meaning, the lump form factor 
receives a 
contribution from $n$ mesons.  In attempting an analogue expansion in the case 
of string field theory, one meets a surprise.  After moving the exponential 
into the kinetic term or equivalently denoting $\, \tilde{T} = \phi\,$
the string theory tachyon 
field equation reads
\be
\left( \left( \partial_x^2 + 1 \right) e^{-c\partial_{x}^{2}} - 2 \right) \phi 
(x) = \bar{g} \phi (x)^2
\ee
with $c = 2\ln \gamma = \ln \, 3^3/4^2$.
   The Ansatz
\be
\phi (x) \sim \phi_0 + a, e^{-\bar{m} x}
\ee
leads to an eigenvalue equation

\be
(\bar{m}^2 + 1 ) e^{-c\bar{m}^2 } - 2 = 0
\ee
for \ $\bar{ m}\,$.  This equation turns out to  have no real solution.  
 With the exponential present the nature of the lump solution has 
changed from that of ordinary field theory.  In particular the decay at 
asymtopic infinity in string field theory has to be stronger than 
a simple exponential.  The non-exponential decay of the full lump (kink) 
solution signals that the form factor will have a more complex physical
meaning.  This feature also necessitates a purely numerical approach to the 
problem which we attempt in the next section.
\medskip
\section{Calculations and Results}
\medskip
The prescence of the higher derivative terms in $(14)$ prevent a 
numerical analysis directly in $x$ space. Upon transforming to momentum 
space, one finds the following nonlinear integral equation
\be
\left( \left( 1-\vec{k}^2 \right) e^{c\vec{k}^{2}} - 2 \right) \phi 
(\vec{k}) = \bar{g} \int d^{25-p}q\phi (\vec{q})\phi (\vec{k}-\vec{q}),
\ee
where $\vec{k}$ is a $25-p$ dimensional vector. 
To solve an integral equation of this type, 
one typically discretizes the problem and solves the resulting matrix equation. In the 
case on hand, the resulting matrix equation is nonlinear and one is forced to search for 
the solution $\phi (\vec{k})$ using a numerical minimization. It is computationally difficult 
and expensive to minimize functions with a large number of variables, and for this reason 
we have worked directly with a spherically symmetric ansatz for the tachyon field. Stable 
numerical solutions were obtained using a lattice having $81$ points. This implies a 
nonlinear minimization problem with $81$ parameters.  The choice of a good 
objective function to be used in the minimization as well as an accurate initial guess are 
crucial to obtain a nontrivial solution. Indeed, in practice, we find that the majority of 
initial guesses lead to the trivial $\phi=0$ solution. Our objective function was 
constructed as follows: Start by making an initial guess, $\hat{G}(k_n)$ for the right 
hand side of $(17)$. This initial guess is used to compute a value for the tachyon
wave function
\be
 \hat{\phi} (k_n) = {\bar{g} \hat{G}(k_n)\over
\left( \left( 1-k_n^2 \right) e^{ck_n^{2}} - 2 \right)}.
\ee
The wave function $\hat{\phi}(k_n)$ can now be used to compute the value of the
left hand side of $(17)$. However, this involves a convolution which must be performed
with care. Evaluating the convolution directly in momentum space is not optimal: 
by exploiting the spherical symmetry of the wave function, the convolution can reduced to
the integration over a single angle and a radius. The integrand contains the factor
$\hat{\phi}(p_n)\hat{\phi}(\delta_n)$ with
\be
\delta_n=|\vec{k}-\vec{p}|=\sqrt{p_n^2+k_m^2-2k_mp_n\cos(\theta_l)}.
\ee
In general, $\delta_n$ does not correpsond to a lattice point and one is forced to  
interpolate from the known points to find $\hat{\phi}(\delta_n)$. A much more efficient 
way of performing the convolution is by transforming to $x$ space, squaring the function 
and then transforming back to $k$ space. By using the spherical symmetry of wave function, 
the Fourier transform can be reduced to a single integration. Thus, the previous double 
integration has been replaced by two single integrations, which is more efficient. In
addition, the integrations only require a knowledge of $\hat{\phi}$ at the lattice points.
After convolving $\hat{\phi}$ with itself to obtain $G(k_n)$, the error, ${\cal E}$ to be 
minimized is constructed as
\be
{\cal E}=\sum_n |G(k_n)-\hat{G}(k_n)|.
\ee
The minimization was performed using the {\tt fmins.m} subroutine of MATLAB, which employs
a simplex search method.  
A suitable initial guess is obtained by taking a function which initially
falls off slightly faster than a Gaussian, but reaches zero at some finite 
momentum. As an example we have shown the wavefunction for the D20 brane
below. 
\newpage

\begin{figure} [h]
\centerline{
\psfig{figure=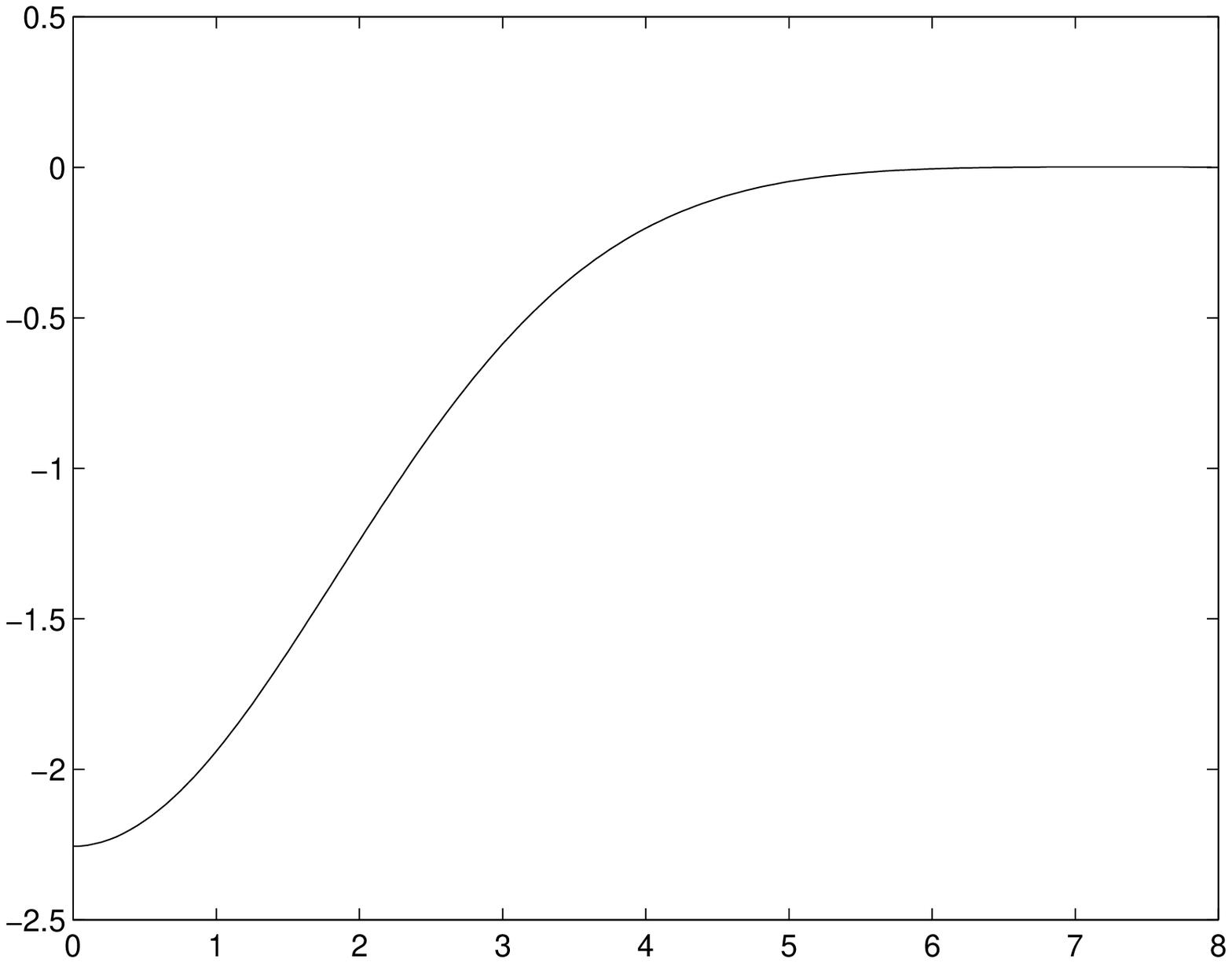,height=10cm,width=10cm}}
\end{figure}

\renewcommand{\baselinestretch}{1}

\begin{quotation}
{\small Fig1: The wave function of the D20 brane as a function of the radial coordinate
$r$. The plot was obtained by taking the Fourier transform of the numerical solution of
equation $(18)$.} 
\end{quotation}

\renewcommand{\baselinestretch}{1.5}

The above form for the wave function is typical. The value of the wave function at the
origin in position space decreased from $\approx -0.62$ for the D24 brane to 
$\approx -4.46 $ for the D18 brane. The point at which the wave function reaches zero
is very nearly constant for all the Dp-branes considered here. In figure 2 we have shown 
the error in the D24 brane solution. The tensions of these
solutions was evaluated directly in momentum space. For example, in the case of the
D24 brane, we compute
\bea \hat{T}_{24}
=2\pi^2 T_{25}&\int& {dp\over 2\pi}\Big({1\over 2}\phi (p)(p^2-1)e^{cp^2}
\phi(-p)+\phi(p)\phi(-p)\nonumber \\
&+& {g\over 3}\int {dk\over 2\pi}\phi (k)\phi (p)
\phi (-p-k)\Big)\nonumber \\ 
={2\pi^2 T_{25}\over 6}&\int& {dp\over 2\pi}\Big(\phi (p)(p^2-1)e^{cp^2}
\phi(-p)+2\phi(p)\phi(-p)\Big).
\eea
\vspace*{.5in}

\begin{figure} [h]
\centerline{
\psfig{figure=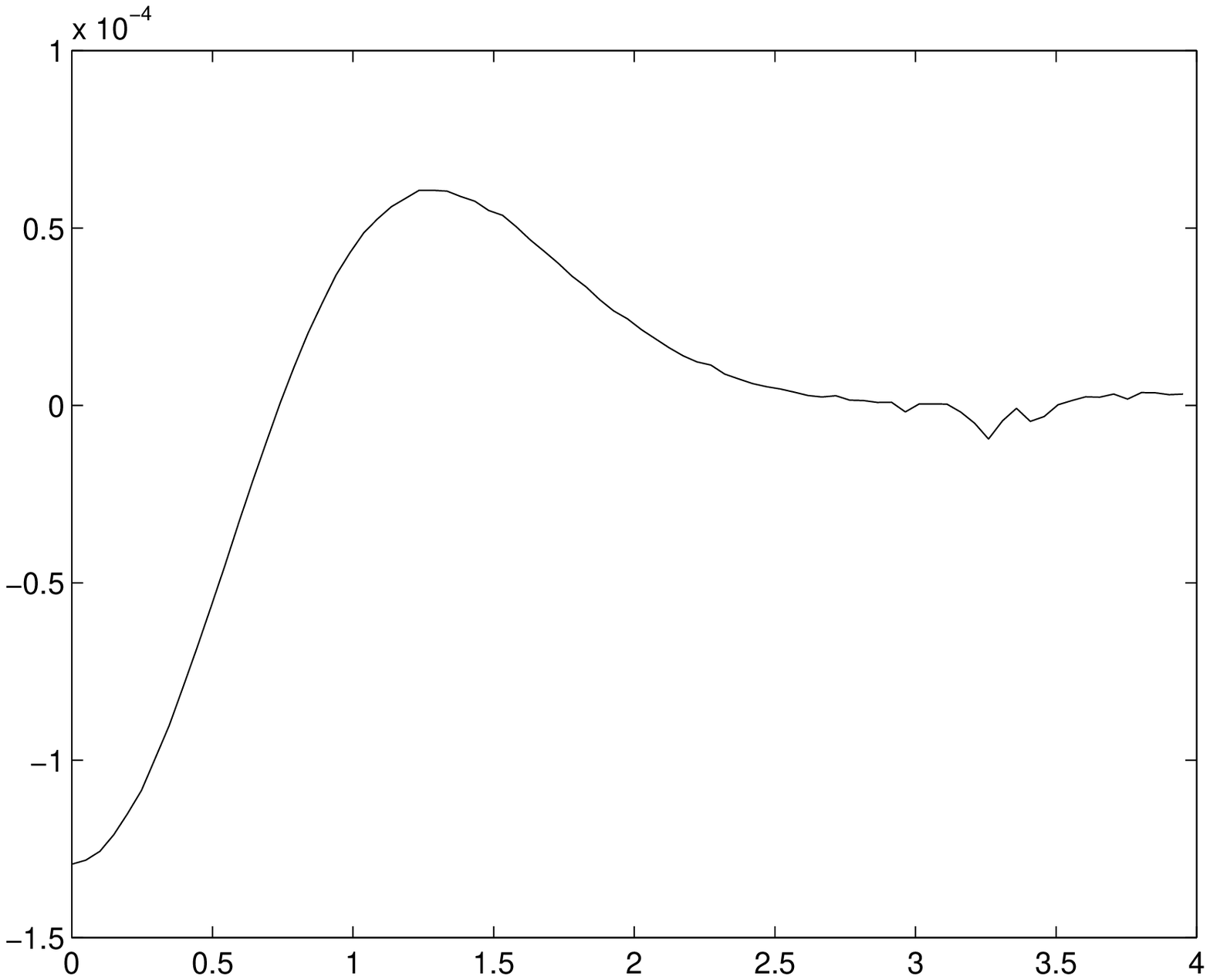,height=10cm,width=10cm}}
\end{figure}

\renewcommand{\baselinestretch}{1}

\begin{quotation}
{\small Fig2: The error in the wave function of the D20 brane as a function of the radial 
coordinate in momentum space $p$. The plot shows ${\cal E}/(\sum_n \hat{G}(k_n)$ as a
function of $k_n$.} 
\end{quotation}

\renewcommand{\baselinestretch}{1.5}

In our conventions, a Dp-brane has tension
\be
T_p=(2\pi)^{25-p}T_{25}.
\ee
In the table below we compare our numerically evaluated tensions, $\hat{T}_p$ with
the above values.
 
\vspace*{.5in}

\begin{center}
\begin{tabular}{|c|c|c|c|}
\hline
{}~~~~~~~~$p$~~~~~~~~ & $\Delta$ $T_p$
~~~~~ &~~~~~~~$\hat{T}_p/T_p$~~~~~~~& ~~~~~~~~$\phi (0)$~~~~~~~~~\\
\hline
24          &   -29.4   &  0.706  & -0.63   \\
\hline
23          &   -27.5   &   0.725  & -0.86   \\
\hline
22          &   -14.3   & 0.857    & -1.19   \\
\hline
21          &   -8.3    &  0.917   & -1.63   \\
\hline
20          &    6.4    &  1.064   & -2.26   \\
\hline
19          &   25.3    &  1.253   & -3.15   \\
\hline
18          &   64.0    &  1.640    & -4.46    \\
\hline
\end{tabular}
\end{center}

\renewcommand{\baselinestretch}{1}
{\small
\normalsize

\begin{center}
{\bf Table I}
\end{center}

\begin{quotation} 
Values of the Dp brane tensions computed using the numerical tachyon lump
solution. The parameter $\Delta T_p$ is defined as
$\Delta T_p ={\hat{T}_p-T_p\over T_{p}}\times 100 .$ $\phi(0)$ is the
value of the wavefunction at the origin in position space. 
\end{quotation}}

\renewcommand{\baselinestretch}{1.5}

Clearly there is no obstacle for the existence of lower p-Brane
solutions. The trend is that the static lump gives systematicaly a growing tension.
This is actually opposite to what one finds in the extreme field theory limit.
Concerning further improvement of present results one expects (esspecialy for lower p) the
relevance of higher massive levels. It is next important to study the direction of their 
contributions. It is also relevant to perform a similar study in superstring 
theory\cite{one, six, seven, sixteen}.


\bigskip

\begin{thebibliography}{**}
\bibitem{one}  A Sen, ``Stable Non-BPS Bound States of BPS D-branes", {\it JHEP} {\bf 9808}:010 (1998), 
het-th/9805019.
\bibitem{two}  A. Sen, ``Descent Relations Among Bosonic D-branes", hep-th/9902105.
\bibitem{three} A. Sen, B. Zwiebach ``Tachyon Condensation in String Field Theory," hep-th/9912249.
\bibitem{four}  V. A. Kostelecky and S. Samuel, ``The Static Tachyon Potential in the Open Bosonic 
String," {\it Phys. Lett.} {\bf B207} (1988) 169.
\bibitem{five}  V. A. Kostelecky and S. Samuel, ``On a Nonperturbative Vacuum for the Open Bosonic String, 
{\it Nucl. Phys.} {\bf B336} (1990) 263-296.
\bibitem{six}  N. Berkovits, ``The Tachyon Potential in Open Neveu-Schwarz String Field Theory," 
hep-th/0001084.
\bibitem{seven}  N. Berkovits, A. Sen and B. Zwiebach, ``Tachyon Condensation in Superstring Field 
Theory," hep-th/0002211
\bibitem{eight}  W. Taylor, ``D-brane Effective Field Theory from String Field Theory," hep-th/0001201.
\bibitem{nine}  N. Moeller and W. Taylor, ``Level Truncation and the Tachyon in Open Bosonic SFT", 
hep-th/0002237.
\bibitem{ten}  E. Witten, ``Non-Commutative Geometry and String Field Theory", {\it Nucl. Phys.} {\bf 
B268}, (1986) 1253.
\bibitem{eleven}  D. J. Gross and A. Jevicki, ``Operator Formulation of Interacting String Field Theory 
(I), (II), (III)" {\it Nucl. Phys.} {\bf B283} (1987) 1; {\it Nucl. Phys.} {\bf B287} (1987) 225; {\it 
Nucl. Phys.} {\bf B293} (1987) 29.
\bibitem{twelve}  E. Cremmer, A. Schwimmer and C. Thorn, ``The vertex function in Witten's Formulation of 
String Field Theory" {\it Phys. Lett.} {\bf B179} (1986) 57.
\bibitem{thirteen}  S. Samuel, ``The Physical and Ghost Vertices in Witten's String Field Theory," {\it 
Phys. Lett.} {\bf B181} (1986) 255.
\bibitem{fourteen}  A. Bogojevic, A. Jevicki and G. Meng, ``Quartic Interactions in Superstring Field 
Theory", Brown preprint, HET-672, 1988 (unpublished) (availabe from KEK preprint library).
\bibitem{fifteen}  J. A. Harvey, P. Kraus ``D-branes as Unstable Lumps in Bosonic Open String Field 
Theory," hep-th/0002117. 
\bibitem{sixteen} T. Yoneya, ``Spontaneously Broken Space-Time Supersymmetry in Open
String Theory without GSO Projection," hep-th/9912255.

\end{thebibliography}
\end{document}